\documentclass[twocolumn,nofootinbib]{revtex4}

\usepackage{blindtext,mathtools,amsmath,amsfonts,amssymb,physics,slashed,esvect,mathrsfs,dsfont,bbold}
\usepackage[utf8]{inputenc}
\usepackage{graphicx}
\usepackage{float}
\usepackage[footnotesize]{caption}

\usepackage{subfig}

\usepackage{multirow}

\usepackage{empheq}

\usepackage{pstricks,pst-node}

\DeclareCaptionJustification{justified}{\leftskip=0pt \rightskip=0pt \parfillskip=0pt plus 1fil}

\usepackage{xcolor}

\mathchardef\Re="023C 
\mathchardef\Im="023D

\pdfoutput=1

\begin{document}
\title{A simplified general relativistic model to analyze the structure of spiral galaxies}

\author{Vincent Deledicque}
\email[]{vincent.deledicq@gmail.com}
\affiliation{No affiliation}

\date{\today}

\begin{abstract}
The dark matter hypothesis, which is not called into question here, explains why typical rotation curves of spiral galaxies do not follow a Keplerian profile. It is however not sufficient in itself to explain why the whole matter distribution in spiral galaxies is such that the rotation curve generally presents a flat profile in the disk region.

To understand this property, a model considering general relativistic effects is developed. It is stressed that the aim is not to explain the flat rotation curve of spiral galaxies without dark matter. More specifically, the analytical stationary solution of an axisymmetric rotating pressureless fluid for the linearized equations of the theory of general relativity is determined. It is demonstrated that this solution leads to some constraints on the rotation curve, by looking to its limit behavior when neglecting general relativistic effects. In particular, the positiveness of the density imposes the rotation curve to be flat in the regions where the density and general relativistic effects are small, such as in the disk region. General relativistic effects  hence remain negligible in the disk region, but their consideration proved to be necessary to establish the aforementioned constraint. Such constraint cannot be derived from a Newtonian approach.

The model is finally applied on two specific cases to demonstrate its ability to predict the rotation curve from a typical density profile along the galactic plane. These examples suggest that for some galaxies, general relativistic effects can be significant close to the bulge region and should be taken into account to have a proper understanding of their rotation curve.
\end{abstract}

\maketitle


\section{Introduction}

When galaxies have been discovered at the beginning of the twentieth century, rapidly the question of their structure has been raised. For spiral galaxies, two main measurements were carried out to answer this question: the luminosity distribution and the rotation curve, i.e., the velocity profile of the stars along the radius of the galactic plane. Comparing the results obtained by both methods led to a still unsolved issue: when applying the classical Newton's laws, the stars in the disk of the galaxies seem to have velocities that are too large with respect to the mass distribution evidenced by the luminosity measurements. While Newton's laws predict, in a first approximation, that stellar rotation velocities in the disk should decrease with distance from the bulge region with an almost Keplerian profile, it is instead observed that they remain constant, up to large distances. Until now, two main distinct ways have been followed to try to solve this mystery.

The first way, initially proposed by \cite{Oort} and \cite{Zwicky}, consists in postulating the existence of significant unseen matter, called dark matter, of which the distribution would be such that it would explain the abnormal high velocities of the stars in the disk region, and this on the basis of a simple application of Newton's laws. Despite the apparent simplicity of this assumption, this leads to postulate the existence of very large quantities of a possibly new kind of matter. 

The second way consists in modifying Newton's laws, in order to avoid postulating the existence of very large quantities of matter which have never directly been observed. A well-known theory proposed in this way is the Modified Newton Dynamics (MOND), established by \cite{Milgrom}.

Recently, a third way has emerged, which consists in using the theory of general relativity from which the Newton's gravitation law is a coarse approximation. Up to now, this way has not been explored very much, and the few tentatives to convince the scientific community of the relevance of this way were unsuccessful. It is indeed largely agreed that the possible effects which would appear by applying the laws of general relativity are so insignificantly small, especially in the disk region, that the Newton's gravitational law is believed appropriate to study the structure of a spiral galaxy. Without being exhaustive, let us cite the following examples. \cite{Cooperstock} used the theory of general relativity to establish in the weak approximation one linear and one nonlinear equation relating the angular velocity to the fluid density. It was then shown on some examples that the rotation curves are consistent with the mass density distributions of the visible matter concentrated in the disk. This article has however been the object of severe criticisms of most of the articles which refer to it (see, for example, \cite{Fuchs} and \cite{Korzynski}). Following this study, \cite{Balasin} studied exact stationary axially symmetric solutions of the 4-dimensional Einstein equations with co-rotating pressureless perfect fluid sources and showed that the Newtonian approach could over-estimate the amount of matter needed to explain rotation curves by more than 30\%. More recently, \cite{Lecore} proposed to explain the rotation curve by taking into account the general relativistic effects developed by large structures of the universe such as clusters of galaxies. 

The first of these three ways is currently the preferred one by most of the scientific community. In this article, we will not call into question the hypothesis of dark matter: dark matter is considered to be necessary to explain the flatness of the rotation curve of spiral galaxies. However, at this stage, there are still several unexplained facts related to this hypothesis. In particular, postulating a distribution of matter for a specific spiral galaxy can help us to understand the rotation curve it exhibits, but it does not explain the physical reason for which matter is distributed as such. Indeed, postulating dark matter explains why the rotation curve does not follow an almost Keplerian profile, but it is not sufficient in itself to explain why the dark matter distribution is such that it generally implies a flat rotation curve in the disk region, and does not follow any other profile. For example, a continuously rising profile of the rotation curve would also require the existence of dark matter, but we don't know why such a profile is not encountered in practice.

In order to have a better understanding of a spiral galaxy's structure, and to answer the previous question, we develop in this article a new model. With respect to the usual approach, which consists in relying on the Newton's law of gravitation, the main difference is the fact that this new model uses the linear approximation of the theory of general relativity. Since the Newton's law of gravitation is a coarser approximation of general relativity than its linear approximation, we may hope that the new model will provide 'better results', in the sense that they could be closer to reality, and that they could provide a better understanding of the physics (such hope is for example illustrated by the case of the precession of the perihelion of Mercury). In our case, we may of course wonder if the Newtonian approximation is not sufficient, and if it is necessary or justified to add the complexity of the linear approximation of general relativity to analyze the structure of spiral galaxies. Will we really learn more by considering general relativistic effects? In this article, a first positive answer will be given to that question. As already stressed, however, the consideration of general relativistic effects will not explain the flatness of the rotation curve without dark matter.

In section $\ref{S1}$ we briefly introduce the linearized approximation of the theory of general relativity. This will constitute the set of equations that we will use in the analysis of section $\ref{S2}$, whose aim is to establish the stationary state of an axisymmetric rotating pressureless fluid. This stationary state will be found to be defined by two differential equations and several algebraic equations. Amongst all the possible mathematical solutions to these equations, only some of them are acceptable on a physical point of view. In section $\ref{S4}$ we identify the most important physical constraints that an acceptable solution must satisfy. We then deduce some structural characteristics of spiral galaxies. In particular, we will demonstrate that the rotation curve of spiral galaxies necessarily tends to a constant profile in the disk region if adequate hypotheses (consistent with the observations or expectations) are made. Finally, in section $\ref{S6}$, we apply the model on two specific cases to demonstrate is ability to predict the rotation curve for a given density profile along the galactic plane.


\section{The linear approximation of general relativity}\label{S1}

In order to analyze the structure of spiral galaxies, we will first establish the stationary state for a rotating fluid assumed to be representative of a spiral galaxy. More details about the hypotheses we will make about this representative rotating fluid are given in section $\ref{S2}$. With this in mind, it appears that we cannot use the classical Newton's law of gravitation to establish the stationary state. If the orbital motion of the stars and other masses can explain the stationary state in the two dimensions of the galactic plane, it is difficult to explain why the galaxy does not collapse in the third (axial) dimension. By considering a pressureless fluid, and in the absence of centrifugal force in this last dimension, the gravitational field as predicted by Newton will indeed attract all the masses towards the symmetry plane of the galaxy. This difficulty disappears if one considers the theory of general relativity. For this last one, the forces which act on the masses present new components which in some cases can produce a repulsive contribution able to counteract the attractive gravitation. It will however not be necessary to use the complete theory of general relativity. For the specific situation we will study, its linear approximation will appear to be sufficient.

The linearized equations of general relativity, whose derivation can be found in several textbooks on general relativity, are obtained by assuming that the metric $g^{\mu\nu}$ can be written as
\begin{equation}
	g^{\mu\nu} = \eta^{\mu\nu} + h^{\mu\nu}\,,
\end{equation}
where $\eta^{\mu\nu}$ is the flat space Minkowski metric and $h^{\mu\nu}$ is such that $|h^{\mu\nu}| \ll 1$. Using the Lorentz gauge, it can be shown that the general relativity equations are then approximated by
\begin{eqnarray}\label{x1}
	\partial_\mu \bar{h}^{\mu\nu} &=& 0\,,
	\\\label{x1bis}
	\Box \bar{h}^{\mu\nu} &=& -2\frac{8\pi G}{c^4}T^{\mu\nu}\,,
\end{eqnarray}
where $G$ is the universal gravitational constant, $c$ is the speed of light, $\textbf{T}$ is the stress-energy tensor and
\begin{eqnarray}
	\bar{h}^{\mu\nu} &=& h^{\mu\nu} - \frac{1}{2}\eta^{\mu\nu}h\,,\nonumber
	\\
	h &=& h^\sigma_{\ \sigma}\,,\nonumber
	\\
	h^\mu_{\ \nu} &=& \eta^{\mu\sigma}h_{\sigma\nu}\,.
\end{eqnarray}
In the approximation of a source term having a low velocity, the stress-energy tensor becomes
\begin{equation}
	T^{00} = \rho c^2\,\,;\,T^{0i} = c\rho v^i\,\,;\,T^{ij} = \rho v^iv^j\,,
\end{equation}
where $\rho$ is the local density, and $v^i$ is the $i^{th}$-component of the fluid velocity. If we consider the stationary case, and if we define the scalar potential $\Phi$ and the vector potential $H^i$ such that
\begin{equation}
	\bar{h}^{00} = \frac{4\Phi}{c^2}\,\,;\,\bar{h}^{0i} = \frac{4H^i}{c}\,\,;\,\bar{h}^{ij} = 0\,,
\end{equation}
the Eq.\ $(\ref{x1})$ and $(\ref{x1bis})$ can be written in the following manner:
\begin{eqnarray}\label{x2}
	\Delta\Phi &=& 4\pi G\rho\,,
	\\\label{x2bis}
	\Delta H^i &=& \frac{4\pi G}{c^2}\rho v^i = \frac{4\pi}{K}\rho v^i\,,
\end{eqnarray}
where we have defined a new constant:
\begin{equation}
	K = \frac{c^2}{G}\,.
\end{equation}
We then define the fields $\mathbf{g}$ and $\mathbf{k}$ as
\begin{eqnarray}
	\mathbf{g} &=& -\mathbf{\nabla}\Phi\,,
	\\
	\mathbf{k} &=& \mathbf{\nabla}\times\mathbf{H}\,.
\end{eqnarray}
The Eq.\ $(\ref{x2})$ and $(\ref{x2bis})$ can then be obtained from the following relations:
\begin{eqnarray}\label{E1}
	\mathbf{\nabla}\times\mathbf{k} &=& -\frac{4\pi}{K}\mathbf{J}\,,
	\\\label{E2}
	\mathbf{\nabla}\cdot\mathbf{k} &=& 0\,,
	\\\label{E3}
	\mathbf{\nabla}\times\mathbf{g} &=& 0\,,
	\\\label{E4}
	\mathbf{\nabla}\cdot\mathbf{g} &=& -4\pi G\rho\,,
\end{eqnarray}
where $\mathbf{J} = \rho \mathbf{v}$. In analogy with electromagnetism which has a similar set of equations, we define $\mathbf{g}$ and $\mathbf{k}$ as being the gravitoelectric and the gravitomagnetic fields, respectively. More details about gravitoelectromagnetism can be found, for example, in \cite{Mashhoon}. Note however that in this reference, the gravitoelectric and gravitomagnetic fields have been defined in a slightly different way.

Also, in the linear approximation of general relativity, the equations of the geodesics become
\begin{equation}\label{E5}
	\frac{d^2\mathbf{x}}{d t^2} = \mathbf{g} + 4\mathbf{v}\times\mathbf{k}\,.
\end{equation}
The aim now is to find a solution to the Eq.\ $(\ref{E1})$ to $(\ref{E5})$ for a rotating fluid representative of a spiral galaxy.


\section{The stationary state of a spiral galaxy}\label{S2}

We will model a galaxy in an idealized manner, using more particularly the following assumptions:
\begin{enumerate}
	\item The galaxy presents a cylindrical symmetry, as well as a symmetry plane through its height, corresponding to the galactic plane. We will therefore use a reference frame expressed in cylindrical coordinates $(r,\theta,z)$ such that its origin lies at the center of the galaxy, the symmetry axis corresponds to $r=0$ and the symmetry plane corresponds to $z = 0$. The basis vectors are denoted $\mathbf{e}_r$, $\mathbf{e}_\theta$ and $\mathbf{e}_z$. The cylindrical symmetry means that the variables such as the density and the velocity do not depend on the polar coordinate $\theta$.
	
	\item The matter inside the galaxy is considered as a pressureless fluid, having a well defined density $\rho$ and velocity $\mathbf{v}$ at each point.
	
	\item The motion of the fluid in the galaxy follows a perfect circle. This means that 
	\begin{equation}
		\mathbf{J} = 0\cdot \mathbf{e}_r + J_\theta\cdot\mathbf{e}_\theta + 0\cdot\mathbf{e}_z\,,
	\end{equation}
	where $J_\theta = \rho v$, and $v$ is the magnitude of the velocity.
\end{enumerate}
On the basis of the assumptions we made, the Eq.\ $(\ref{E1})$ to $(\ref{E4})$ can thus be written as
\begin{eqnarray}\label{A1}
	&&\frac{\partial k_\theta}{\partial z} = 0\,,
	\\\label{A2}
	&&\frac{\partial k_r}{\partial z} - \frac{\partial k_z}{\partial r} = -\frac{4\pi}{K}\rho v\,,
	\\\label{A3}
	&&\frac{\partial(rk_\theta)}{\partial r} = 0\,,
	\\\label{A4}
	&&\frac{\partial k_r}{\partial r} + \frac{k_r}{r} + \frac{\partial k_z}{\partial z} = 0\,,
	\\\label{A5}
	&&\frac{\partial g_\theta}{\partial z} = 0\,,
	\\\label{A6}
	&&\frac{\partial g_r}{\partial z} - \frac{\partial g_z}{\partial r} = 0\,,
	\\\label{A7}
	&&\frac{\partial (rg_\theta)}{\partial r} = 0\,,
	\\\label{A8}
	&&\frac{\partial g_r}{\partial r} + \frac{g_r}{r} + \frac{\partial g_z}{\partial z} = -4\pi G\rho\,,
\end{eqnarray}
where we used the notation $\mathbf{g} = (g_r, g_\theta, g_z)$ and $\mathbf{k} = (k_r, k_\theta, k_z)$. Also, Eq.\ $(\ref{E5})$ becomes
\begin{eqnarray}\label{A9}
	-\frac{v^2}{r} &=& g_r + 4v k_z\,,
	\\\label{A10}
	0 &=& g_\theta\,,
	\\\label{A11}
	0 &=& g_z - 4v k_r\,.
\end{eqnarray}
We thus have a set of 11 equations (8 differential equations and 3 algebraic equations) for the unknowns $k_\theta$, $k_r$, $k_z$, $g_\theta$, $g_r$, $g_z$, $v$ and $\rho$.

It is important to emphasize that the Eq.\ $(\ref{A1})$ to $(\ref{A8})$ are valid in the whole space, whereas the Eq.\ $(\ref{A9})$ to $(\ref{A11})$ are only valid in the regions where an equilibrium state can exist. Inside the galaxy, an equilibrium state has necessarily been reached (otherwise masses could not be present there), but outside of it, such an equilibrium state will generally not exist. This is explained by the fact that outside of the galaxy, where the density is zero, the equilibrium state is determined by an over-constrained system of equations (by imposing $\rho=0$ we lost a degree of freedom). Indeed, the velocity has to verify two equations, namely the Eq.\ $(\ref{A9})$ and $(\ref{A11})$, in which the gravitoelectromagnetic fields have, at some point, values that are fixed for a specific galaxy. Given these values, the Eq.\ $(\ref{A9})$ and $(\ref{A11})$ will in general admit no solution for $v$, except at very specific locations. The relations we will develop below are thus only valid inside the galaxy, and not elsewhere. 

\subsection{Derivation of the gravitoelectromagnetic fields}

From Eq.\ $(\ref{A1})$ we deduce that $k_\theta$ does not depend on $z$, and from Eq.\ $(\ref{A3})$ we find that
\begin{equation}
	k_\theta = \frac{k_1}{r}\,,
\end{equation}
where $k_1$ is a constant of integration. For physical reasons however ($k_\theta$ can only be zero at $r = 0$), $k_1$ is necessarily zero, and $k_\theta = 0$ everywhere.

From Eq.\ $(\ref{A11})$ we deduce that
\begin{eqnarray}\label{A15}
	\frac{\partial g_z}{\partial r} &=& 4\frac{\partial v}{\partial r}k_r + 4v\frac{\partial k_r}{\partial r}\,,
	\\\label{A16}
	\frac{\partial g_z}{\partial z} &=& 4\frac{\partial v}{\partial z}k_r + 4v\frac{\partial k_r}{\partial z}\,.
\end{eqnarray}
In a similar way, from Eq.\ $(\ref{A9})$ we have
\begin{equation}\label{A9A}
	g_r = -\frac{v^2}{r} - 4v k_z\,,
\end{equation}
which implies
\begin{eqnarray}\label{A18}
	\frac{\partial g_r}{\partial r} &=& \frac{v^2}{r^2} - 2\frac{v}{r}\frac{\partial v}{\partial r} - 4\frac{\partial v}{\partial r}k_z - 4v\frac{\partial k_z}{\partial r}\,,
	\\\label{A19}
	\frac{\partial g_r}{\partial z} &=& -2\frac{v}{r}\frac{\partial v}{\partial z} - 4\frac{\partial v}{\partial z}k_z - 4v\frac{\partial k_z}{\partial z}\,.
\end{eqnarray}
Firstly, by inserting the Eq.\ $(\ref{A15})$ and $(\ref{A19})$ in Eq.\ $(\ref{A6})$ we obtain
\begin{equation}\label{A20}
	\frac{v}{r}\frac{\partial v}{\partial z} + 2\frac{\partial v}{\partial z}k_z + 2v\left(\frac{\partial k_r}{\partial r} + \frac{\partial k_z}{\partial z}\right) + 2\frac{\partial v}{\partial r}k_r = 0\,.
\end{equation}
Using then Eq.\ $(\ref{A4})$ to replace the terms in parentheses, we may write Eq.\ $(\ref{A20})$ as
\begin{equation}\label{A21A}
	\frac{v}{r}\frac{\partial v}{\partial z} + 2\frac{\partial v}{\partial z}k_z + 2k_r\left(\frac{\partial v}{\partial r} - \frac{v}{r}\right) = 0\,.
\end{equation}
Let us multiply this relation by $\left(\partial v/\partial r + v/r\right)$:
\begin{eqnarray}\label{A22}
	&&\frac{v}{r}\frac{\partial v}{\partial z}\left(\frac{\partial v}{\partial r} + \frac{v}{r}\right) + 2\frac{\partial v}{\partial z}k_z\left(\frac{\partial v}{\partial r} + \frac{v}{r}\right)\nonumber
	\\
	&& + 2k_r\left(\left(\frac{\partial v}{\partial r}\right)^2 - \left(\frac{v}{r}\right)^2\right) = 0\,.
\end{eqnarray}
Secondly, by inserting the Eq.\ $(\ref{A16})$, $(\ref{A9A})$ and $(\ref{A18})$ in Eq.\ $(\ref{A8})$ we obtain
\begin{eqnarray}\label{A22B}
	&&-\frac{v}{r}\frac{\partial v}{\partial r} - 2k_z\left(\frac{\partial v}{\partial r} + \frac{v}{r}\right) + 2v\left(\frac{\partial k_r}{\partial z} - \frac{\partial k_z}{\partial r}\right)\nonumber
	\\
	&& + 2\frac{\partial v}{\partial z}k_r = -2\pi G\rho\,.
\end{eqnarray}
Using Eq.\ $(\ref{A2})$ in order to replace $\left(\partial k_r/\partial z - \partial k_z/\partial r\right)$, this can be simplified in the form
\begin{eqnarray}\label{A22C}
	&&-\frac{v}{r}\frac{\partial v}{\partial r} - 2k_z\left(\frac{\partial v}{\partial r} + \frac{v}{r}\right) + 2\frac{\partial v}{\partial z}k_r\nonumber
	\\
	&& = -\frac{2\pi\rho}{K}\left(KG - 4v^2\right)\,.
\end{eqnarray}
It will be useful to multiply this relation by $\partial v/\partial z$:
\begin{eqnarray}\label{A23}
	&&-\frac{v}{r}\frac{\partial v}{\partial r}\frac{\partial v}{\partial z} - 2k_z\left(\frac{\partial v}{\partial r} + \frac{v}{r}\right)\frac{\partial v}{\partial z} + 2\left(\frac{\partial v}{\partial z}\right)^2k_r\nonumber
	\\
	&& = -\frac{2\pi\rho}{K}\left(KG - 4v^2\right)\frac{\partial v}{\partial z}\,.
\end{eqnarray}
If we then sum the Eq.\ $(\ref{A22})$ and $(\ref{A23})$, and isolate $k_r$, we obtain
\begin{equation}\label{A25}
	k_r = \frac{1}{2}\frac{\partial v}{\partial z}\left(\frac{\frac{2\pi\rho}{K}\left(KG - 4v^2\right) + \left(\frac{v}{r}\right)^2}{\left(\frac{v}{r}\right)^2 - \left(\frac{\partial v}{\partial r}\right)^2 - \left(\frac{\partial v}{\partial z}\right)^2}\right)\,.
\end{equation}
If $\partial v/\partial z \neq 0$, by using Eq.\ $(\ref{A25})$ in Eq.\ $(\ref{A21A})$ we deduce that
\begin{equation}\label{A26}
	k_z = -\frac{1}{2}\frac{v}{r} - \frac{1}{2}\left(\frac{\partial v}{\partial r} - \frac{v}{r}\right)\left(\frac{\frac{2\pi\rho}{K}\left(KG - 4v^2\right) + \left(\frac{v}{r}\right)^2}{\left(\frac{v}{r}\right)^2 - \left(\frac{\partial v}{\partial r}\right)^2 - \left(\frac{\partial v}{\partial z}\right)^2}\right)\,.
\end{equation}
In fact, using Eq.\ $(\ref{A22C})$, we may convince us that Eq.\ $(\ref{A26})$ is valid even if $\partial v/\partial z = 0$.

We have been able to express $k_r$ and $k_z$ in function of $\rho$, $v$ and its derivatives. It can be shown in an equivalent way that $g_\theta = 0$ and that we have very similar expressions for $g_r$ and $g_z$. More particularly, we have:
\begin{equation}\label{A44}
	g_z = 2v\frac{\partial v}{\partial z}\left(\frac{\frac{2\pi\rho}{K}\left(KG - 4v^2\right) + \left(\frac{v}{r}\right)^2}{\left(\frac{v}{r}\right)^2 - \left(\frac{\partial v}{\partial r}\right)^2 - \left(\frac{\partial v}{\partial z}\right)^2}\right)
\end{equation}
and
\begin{equation}\label{46A}
	g_r = \frac{v^2}{r} - 2v\left(\frac{v}{r} - \frac{\partial v}{\partial r}\right)\left(\frac{\frac{2\pi\rho}{K}\left(KG - 4v^2\right) + \left(\frac{v}{r}\right)^2}{\left(\frac{v}{r}\right)^2 - \left(\frac{\partial v}{\partial r}\right)^2 - \left(\frac{\partial v}{\partial z}\right)^2}\right)\,.
\end{equation}

\subsection{The density field}

For practical reasons, we define the non-dimensional variable $F$ as being
\begin{equation}\label{F}
	F = \frac{\frac{2\pi\rho}{K}\left(KG - 4v^2\right) + \left(\frac{v}{r}\right)^2}{\left(\frac{v}{r}\right)^2 - \left(\frac{\partial v}{\partial r}\right)^2 - \left(\frac{\partial v}{\partial z}\right)^2}\,,
\end{equation}
We may then write the expressions of the fields $k_r$, $k_z$, $g_r$ and $g_z$ established above in the following form:
\begin{eqnarray}\label{111}
	k_r &=& \frac{1}{2}\frac{\partial v}{\partial z}F\,,
	\\\label{112}
	k_z &=& -\frac{1}{2}\frac{v}{r} - \frac{1}{2}\left(\frac{\partial v}{\partial r} - \frac{v}{r}\right)F\,,
	\\\label{113}
	g_z &=& 2v\frac{\partial v}{\partial z}F\,,
	\\\label{114}
	g_r &=& \frac{v^2}{r} + 2v\left(\frac{\partial v}{\partial r} - \frac{v}{r}\right)F\,.
\end{eqnarray}
The density $\rho$ is obtained by inverting Eq.\ $(\ref{F})$:
\begin{equation}\label{sdf}
	\rho = \frac{K}{2\pi}\frac{F\left(\left(\frac{v}{r}\right)^2 - \left(\frac{\partial v}{\partial r}\right)^2 - \left(\frac{\partial v}{\partial z}\right)^2\right) - \left(\frac{v}{r}\right)^2}{KG - 4v^2}\,.
\end{equation}
Hence, the fields $k_r$, $k_z$, $g_r$, $g_z$ and $\rho$ are expressed now in terms of $F$ and $v$.

\subsection{The velocity field and the $F$-field}

From the 11 independent equations $(\ref{A1})$ to $(\ref{A11})$, 4 have been used to establish the gravitomagnetic field, and 5 have been used to establish the gravitoelectric field (in fact for this latter one, 3 equations were really useful, and 2 equations were redundant). Hence, we still have 2 independent equations for the fields $F$ and $v$.

Taking the derivatives of $k_r$ and $k_z$ with respect to $r$ and $z$ in the Eq.\ $(\ref{111})$ and $(\ref{112})$, and inserting the results in Eq.\ $(\ref{A4})$, we obtain
\begin{equation}\label{AA22}
	\frac{1}{r}\frac{\partial v}{\partial z}\left(2Fr + r^2\frac{\partial F}{\partial r} - r\right) = r^2\frac{\partial}{\partial r}\left(\frac{ v}{r}\right)\frac{\partial F}{\partial z}\,.
\end{equation}	
Then, still using the expressions of the derivatives of $k_r$ and $k_z$ with respect to $r$ and $z$ obtained from the Eq.\ $(\ref{111})$ and $(\ref{112})$, and inserting them in Eq.\ $(\ref{A2})$, we get
\begin{eqnarray}\label{eqeq}
	&&\frac{\partial}{\partial z}\left(\frac{1}{2}\frac{\partial v}{\partial z}F\right) + \frac{\partial}{\partial r}\left[\frac{1}{2}\frac{v}{r} + \frac{1}{2}\left(\frac{\partial v}{\partial r} - \frac{v}{r}\right)F\right]\nonumber
	\\
	&& = 2v\frac{\left(\frac{v}{r}\right)^2 - F\left(\left(\frac{v}{r}\right)^2 - \left(\frac{\partial v}{\partial r}\right)^2 - \left(\frac{\partial v}{\partial z}\right)^2\right)}{KG - 4v^2}\,.
\end{eqnarray}
This is a complex elliptical non linear partial differential equation for $v$. If we could define appropriate boundary conditions for $v$ and $F$, we would be able to solve these two equations, and afterwards, completely determine the fields $\rho$, $k_r$, $k_z$, $g_r$ and $g_z$. This is out of the scope of this article.


\section{Qualitative analysis of the structure of a spiral galaxy}\label{S4}

We have developed the relations that must be verified by a stationary rotating pressureless fluid. Several mathematical solutions could exist to these relations, but all are not necessarily acceptable, because some physical constraints have to be verified. In particular, we expect that spiral galaxies are submitted to the two following constraints.

Firstly, in the upper half of the domain $(z>0)$, we expect that the $z$ component of the gravitoelectric field points towards the symmetry plane of the galaxy, in other words that $g_z<0$. By symmetry we expect that $g_z>0$ in the lower half of the domain $(z<0)$. On the symmetry plane we have $g_z = 0$. If we consider the case for which $v > 0$, Eq. $(\ref{111})$ and $(\ref{113})$ show that $k_r$ has the same sign as $g_z$. So in the upper half of the domain we have $k_r < 0$ whereas in the lower half we have $k_r > 0$.

Secondly, we expect the density to be everywhere positive. This assumption could seem obvious, but if it is admitted that some unknown matter (called dark matter) could explain the rotation curve, we may not exclude theoretically that other kinds of matter, having negative masses, could also exist. Negative masses have already been studied in the past (see \cite{Bondi}), and it was concluded that the theory of general relativity does not exclude it, even if it leads to negative energies. We will not consider such a possibility in this analysis, and we will impose the density to be positive.

We have typically $4v^2 < KG = c^2$. For example, the maximum value of the velocity in the Milky Way is about $250$ $km\, s^{-1}$, and for this case $4v^2/c^2 \approx 2.8\times 10^{-6}$. Since $4v^2 < KG$, the denominator of Eq.\ $(\ref{sdf})$ is positive, and in order to have a positive density, the numerator should also be positive:
\begin{equation}\label{aze78bis}
	F\left(\left(\frac{v}{r}\right)^2 - \left(\frac{\partial v}{\partial r}\right)^2 - \left(\frac{\partial v}{\partial z}\right)^2\right) - \left(\frac{v}{r}\right)^2 > 0\,.
\end{equation}

Form these two physical constraints, we will now deduce some characteristics of a typical spiral galaxy's structure. With this in mind, it will be helpful to sketch a picture of the gravitomagnetic field that would be expected for a spiral galaxy. 

\begin{figure}
	\centering\includegraphics[width=7cm]{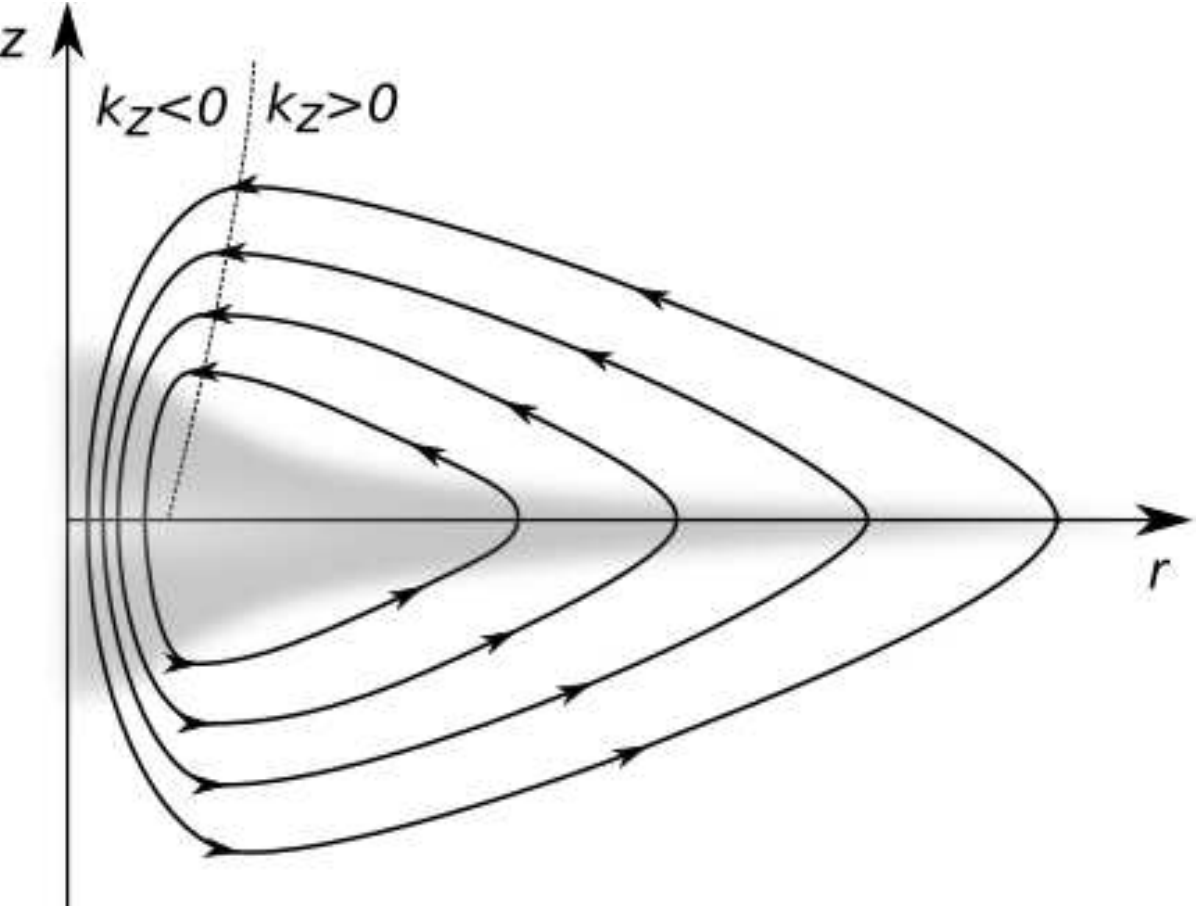}
	\caption{Qualitative illustration of the current lines of the gravitomagnetic field developed by a spiral galaxy.}\label{Galaxie1}
\end{figure}

By analogy with electromagnetism, whose field equations are identical at the exception of the physical constants, a spiral galaxy can be considered as an assembly of successive concentric current loops having different radii, in which the masses are the equivalent of the electric charges. Conceptually, the gravitomagnetic field developed by a spiral galaxy will not be very different from the one of the classical magnetic field around a current loop, albeit a little bit different in its shape due to the fact that the galaxy is more complex than a single current loop. Typical current lines of the gravitomagnetic field developed by a spiral galaxy are qualitatively illustrated in figure $\ref{Galaxie1}$. Due to the vanishing divergence of $\textbf{k}$, the current lines of this field must be closed, as for the classical magnetic field. Since, as shown above for the case $v>0$, we have $k_r < 0$ in the upper half of the domain (and inversely $k_r > 0$ in the lower half), we know that these current lines turn in an anticlockwise way. Note that the gravitomagnetic field could be a little bit more complex than the one sketched in figure $\ref{Galaxie1}$, especially for intermediate radii, but the important point is that typically, $k_z<0$ in a region around the symmetry axis, while $k_z>0$ in the farthest region from the symmetry axis.

We expect thus to have $k_z>0$ for large radii, and, in particular, in a more or less large part of the disk region. Considering Eq.\ $(\ref{A9})$, we deduce that in this latter region, the gravitomagnetic field leads to lower the rotation velocity with respect to what it would be in the presence of the single classical gravitational field, indicating that it certainly cannot explain its flatness (with respect to a Keplerian profile, the gravitomagnetic field would have to increase the rotation velocity). On the other hand, in the central region of the bulge, we have $k_z<0$, meaning that the gravitomagnetic field leads to increase the rotation velocity with respect to what it would be in the presence of the single classical gravitational field.

\begin{figure}
	\centering\includegraphics[width=7cm]{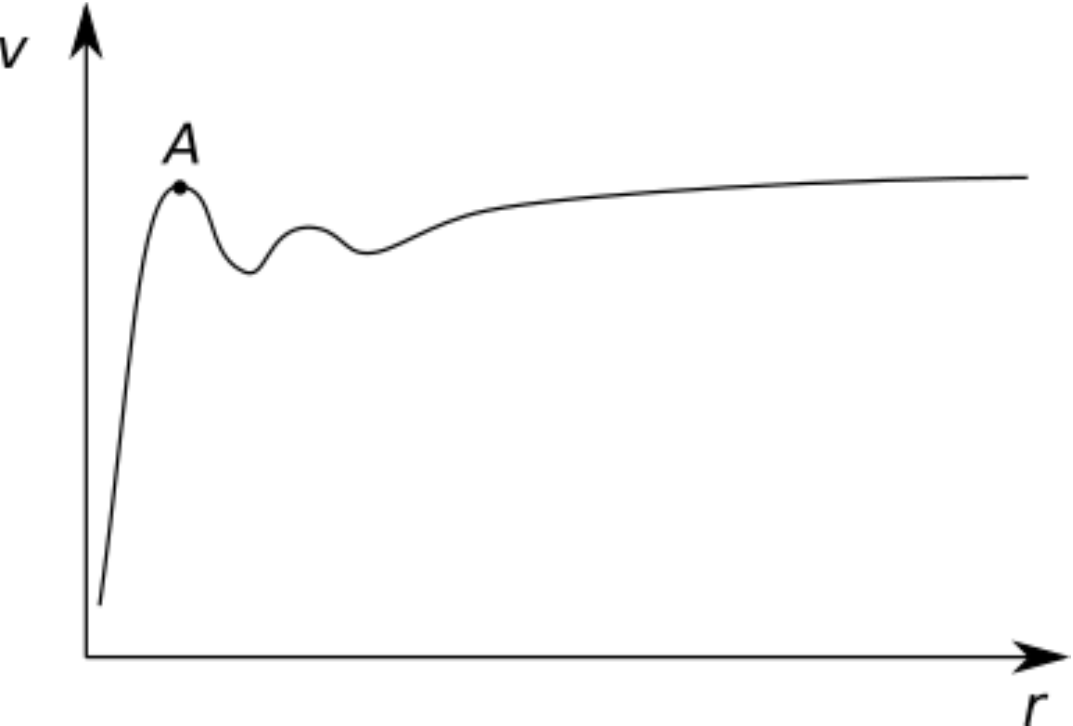}
	\caption{Typical rotation curve of a spiral galaxy.}\label{Galaxie3}
\end{figure}

On the basis of some typical rotation curves, we may convince us that the region in which $k_z<0$ is very narrow with respect to the characteristic dimensions of a spiral galaxy. Let us indeed consider the rotation curve illustrated in figure $\ref{Galaxie3}$, which is qualitatively representative of typical rotation curves. On this figure, point $A$ is characterized by the fact that $\partial v/\partial r = 0$, and on the galactic plane where $g_z = 0$ we also have $\partial v/\partial z = 0$. The positiveness of the density (see Eq.\ $\ref{aze78bis}$) then implies that $F>1$. Using also Eq.\ $(\ref{112})$, we show that at this point $k_z = \left(F - 1\right)v/r$. Since $F > 1$, this means that $k_z>0$ at point $A$, and thus that the region in which $k_z < 0$ must lie closer to the origin. 

Concerning the orders of magnitude, we expect large values of $k_r$ in the bulge. This is justified by the fact that the large densities observed in this region suggest us that we have a significant gravitoelectric field $g_z$, and in order to avoid the collapse in the axial direction, Eq.\ $(\ref{A11})$ requires therefore a significant gravitomagnetic field $k_r$, especially if the velocity is small. On the other hand, in the disk region, we expect a much smaller gravitoelectric field $g_z$, and Eq.\ $(\ref{A11})$ requires at equilibrium a smaller gravitomagnetic field $k_r$ as well.

Also, we may probably not neglect $k_z$ in the region around the symmetry axis. Indeed, as already sketched, a spiral galaxy can be considered as an assembly of successive concentric current loops. In some places, these current loops develop gravitomagnetic fields which for some of them combine in a destructive manner (meaning that they cancel each other), and for the others combine in a constructive manner (meaning that they enforce each other). But the more we approach the center of the galaxy, the more these current loops combine in an overall constructive manner. In particular, in the very central region, $k_z$ will probably reach significant values. This is justified by the fact that, as illustrated above, the region in which $k_z<0$ is probably very narrow. Since the flux trough this region has to be identical in magnitude to the flux trough the whole region where $k_z>0$, which expands over quite large distances, we may expect that the flux density be large in the former region, hence that $k_z$ be significant there.

On the other hand, we do not expect significant values of $k_z$ in the disk region. This assumption will moreover be justified by the fact that it is part of the explanation of the flatness of the rotation curve, as we will now see.

In a rough approximation, we generally represent a spiral galaxy as a central bulge region in which the density is quite high, and a disk region, in which we assume that the density is negligible. With such a configuration, the application of the classical Newton's laws implies that the velocity presents a Keplerian profile in the disk region, at least if we suppose that the contribution of the masses outside of the bulge to the gravitational field are insignificant. We demonstrate here that the general relativistic approach leads to another conclusion, i.e., that the velocity profile necessarily presents a constant profile, in accordance with the observations. As already generally admitted, this then requires the existence of larger quantities of matter than observed, confirming thus the existence of the halo. This demonstration will be given on the galactic plane, where $\partial v/\partial z = 0$.

As explained above, we expect the gravitomagnetic field to be negligible in the disk region. Regarding Eq.\ $(\ref{A9})$, the smallness of $k_z$ means that the last term is small in comparison with the two other terms. In particular:
\begin{equation}
	| k_z | \ll \frac{1}{4}\frac{v}{r}\,.
\end{equation}
In this case, Eq.\ $(\ref{A9})$ tends to the classical Newtonian law:
\begin{equation}\label{wiw}
	g_r \approx -\frac{v^2}{r}\,.
\end{equation}

The fact that the density is negligible in the disk region implies that the local masses do not influence significantly the local gravitoelectric field, and thus that the term related to the density in Eq.\ $(\ref{A8})$ is small in comparison with $g_r$:
\begin{equation}\label{ertert}
	4\pi G \rho r \ll | g_r |\,.
\end{equation}
Taking into account Eq.\ $(\ref{wiw})$ this latter inequality becomes
\begin{equation}\label{dfgdfg}
	\pi G \rho \ll \frac{1}{4}\frac{v^2}{r^2}\,.
\end{equation}

Let us now examine the expression of the density, Eq.\ $(\ref{sdf})$. Since typically $KG = c^2 \gg 4v^2$, we will admit that $KG - 4v^2 \approx c^2$. Reminding that we consider the galactic plane ($\partial v/\partial z = 0$), we may then write
\begin{equation}
	\rho\pi G \approx \frac{1}{2}F\left(\left(\frac{v}{r}\right)^2 - \left(\frac{\partial v}{\partial r}\right)^2\right) - \frac{v^2}{2r^2}\,.
\end{equation}
Considering also Eq.\ $(\ref{112})$, this last relation can be written as
\begin{eqnarray}\label{apa}
	&&\rho\pi G \approx \left(\frac{1}{2}\frac{v}{r} + k_z\right)\left(\frac{v}{r} + \frac{\partial v}{\partial r}\right) - \frac{v^2}{2r^2}\nonumber
	\\
	&& = \frac{\partial v}{\partial r}\left(\frac{1}{2}\frac{v}{r} + k_z\right) + k_z\frac{v}{r}\,.
\end{eqnarray}
Using this expression in the inequality $(\ref{dfgdfg})$ leads to
\begin{equation}
	\frac{\partial v}{\partial r}\left(\frac{1}{2}\frac{v}{r} + k_z\right) + k_z\frac{v}{r} \ll \frac{1}{4}\frac{v^2}{r^2}\,.
\end{equation}
Since $| k_z | \ll v/(4r) < v/(2r)$, we deduce that
\begin{equation}
	\frac{\partial v}{\partial r} \ll \frac{1}{2}\frac{v}{r}\,.
\end{equation}
We may thus neglect $\partial v/\partial r$ in front of $v/r$. Doing so in Eq.\ $(\ref{112})$ and neglecting here also $k_z$ in front of $v/(2r)$, we conclude that in the disk region $F \approx 1$. With this value of $F$, the constraint $(\ref{aze78bis})$ on the density imposes then that $\partial v/\partial r$ vanishes:
\begin{equation}
	\frac{\partial v}{\partial r} \approx 0\,.
\end{equation}
A vanishing density field and a negligible gravitomagnetic field are necessarily associated with an almost constant velocity profile along the radius. In the disk region, the velocity will thus keep the value it has reached at the border of the bulge. This indicates that the model we have developed on the basis of the linearized equations of general relativity is able to explain the reason for which the rotation curve of spiral galaxies systematically reaches a constant profile in the disk region. How this flat profile is reached is explained by the dark matter hypothesis.

Note finally that we may also reason in the opposite way: a constant velocity profile and the assumption of a negligible gravitomagnetic field are necessarily associated with a vanishing density field. Indeed, on the basis of Eq.\ $(\ref{112})$ these assumptions lead to $F \approx 1$, which then implies according to Eq.\ $(\ref{sdf})$ that the density vanishes.


\section{Application}\label{S6}

In order to illustrate the ability of the model to predict the rotation curve for a given density profile along the galactic plane, we apply it on two specific cases. Therefore, from Eq.\ $(\ref{apa})$, we note that the density can be written as
\begin{equation}\label{rhonouveau}
	\rho = \frac{1}{2\pi G}\left(\frac{v}{r}\frac{\partial v}{\partial r} + 2k_z\left(\frac{v}{r}+\frac{\partial v}{\partial r}\right)\right)\,.
\end{equation}

In a first simplified approach, let us assume that $k_z$ is negligibly small: $k_z \rightarrow 0$. Then Eq.\ $(\ref{rhonouveau})$ simplifies as
\begin{equation}\label{yue}
	\rho = \frac{1}{2\pi G}\frac{v}{r}\frac{\partial v}{\partial r}\,.
\end{equation}
To solve this differential equation for $v$, we need to describe the density in function of $r$. It is well-known that in the disk region, the density presents typically a decreasing exponential profile:
\begin{equation}\label{rho}
	\rho = \rho_0 e^{-\alpha_0 r}\,,
\end{equation} 
where $\rho_0$ and $\alpha_0 $ are constants. Still in a first simplified approach, we will assume that this profile is representative of the galaxy's density on the whole galactic plane.

So, using the expression of $\rho$ in Eq.\ $(\ref{yue})$, we obtain a differential equation for $v$ which can be solved analytically:
\begin{equation}\label{ddd}
	v(r) = \frac{1}{\alpha_0}\sqrt{\alpha_0^2v_\infty^2 - 4\pi \rho_0 G\left(\alpha_0 r + 1\right)e^{-\alpha_0 r}}\,,
\end{equation}
where $v_\infty$ is a constant. Let us illustrate that this simple relation is able to describe the rotation curve of some typical spiral galaxies, in particular the ones characterized by a shoulder type rotation curve. To do so, we need to fix the parameters $\rho_0$, $\alpha_0$ and $v_\infty$. For a given spiral galaxy, these parameters can be obtained as follows:
\begin{enumerate}
	\item When $r \rightarrow \infty$, we deduce from Eq.\ $(\ref{ddd})$ that $v \rightarrow v_\infty$. Hence, $v_\infty$ corresponds to the limit value to which to velocity profile tends.
	
	\item We impose the derivative of the velocity profile to tend to its specific value when $r \rightarrow 0$, and we impose that $v \rightarrow 0$ when $r \rightarrow 0$. Since
	\begin{equation}
		\frac{\partial v}{\partial r} = 2\pi G\rho_0e^{-\alpha_0 r}\frac{r}{v}\,,
	\end{equation}
	and using L'H\^opital's rule to evaluate this expression when $r \rightarrow 0$, we find
	\begin{equation}
		\frac{\partial v}{\partial r}|_{r\rightarrow 0} \rightarrow \sqrt{2\pi G\rho_0}\,.
	\end{equation}
	Hence $\rho_0$ can be fixed from the slope of the rotation curve near the center of the galaxy.
	
	\item Still using the assumption that $v \rightarrow 0$ when $r \rightarrow 0$, we get
	\begin{equation}
		\alpha_0^2v_\infty^2 - 4\pi \rho_0G = 0\,,
	\end{equation}
	from which $\alpha_0$ can be deduced.
\end{enumerate}

\begin{figure}
\centering
\includegraphics[width=\linewidth]{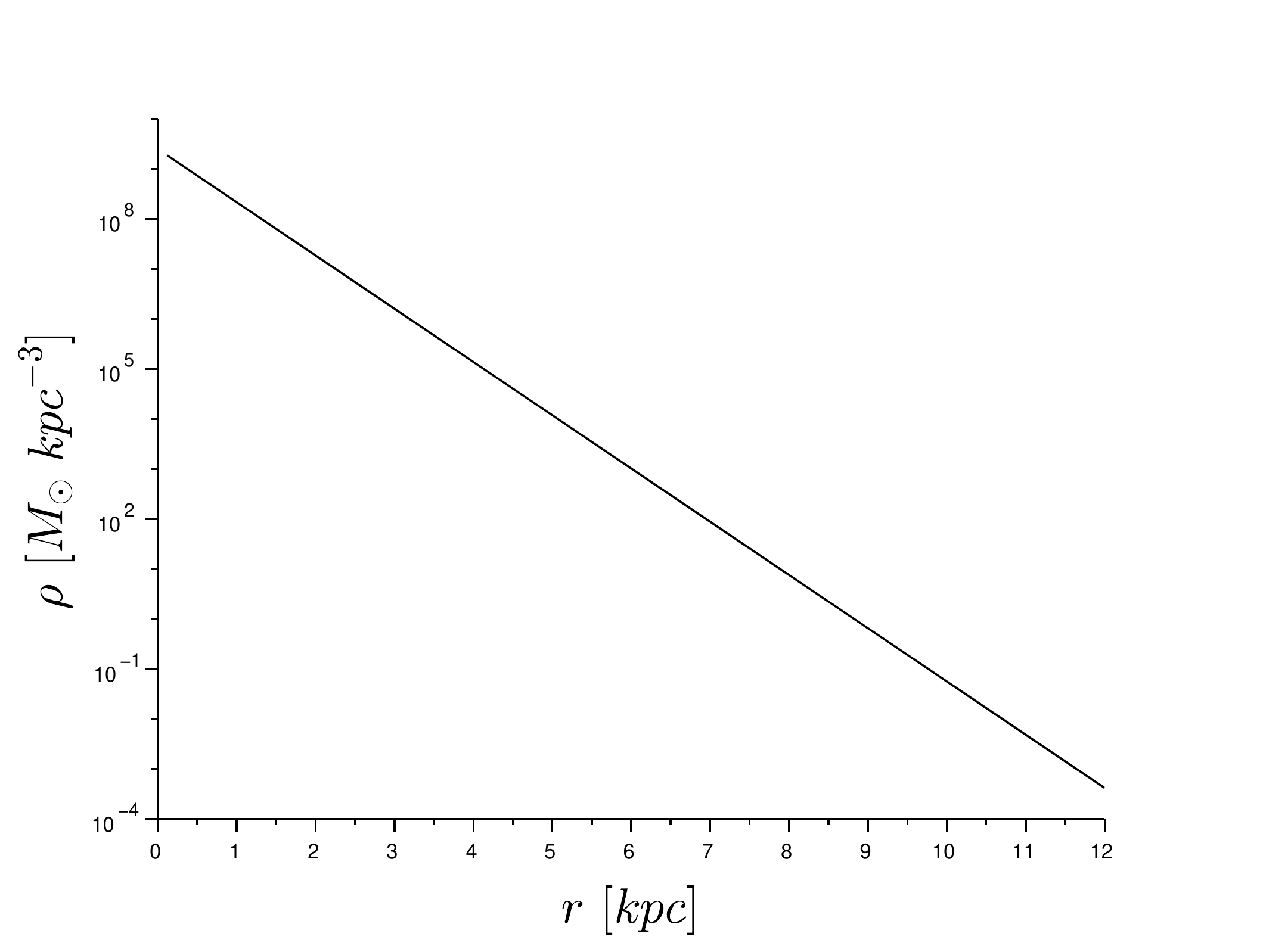}
\captionsetup{width=0.5\textwidth}\caption{Theoretical density profile along the galactic plane of spiral galaxy NGC 3198.}\label{Densite}
\end{figure}

\begin{figure}
	\centering
	\includegraphics[width=\linewidth]{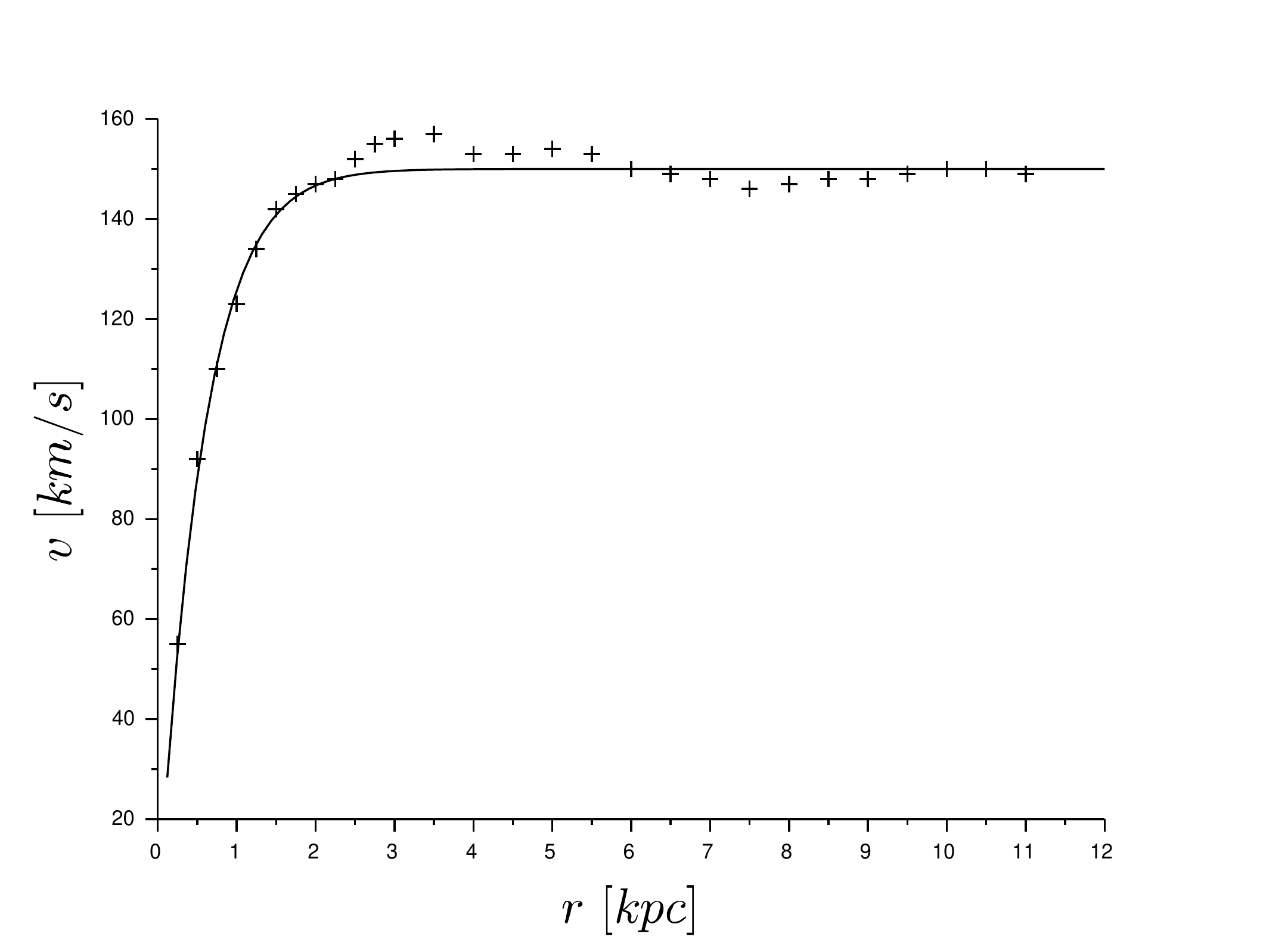}
	\captionsetup{width=0.5\textwidth}\caption{Rotation curve of spiral galaxy NGC 3198. The solid line represents the prediction of the model, while de $+$ marks are observational data from \cite{Begeman}.}\label{Vitesse}
\end{figure}

The validity of the model can be illustrated on the following example. We consider the spiral galaxy NCG 3198, and use for the real rotation curve the data from \cite{Begeman}. We find for this case $\rho_0 = 1.69\times 10^{-19} \ kg\ m^{-3}$, $\alpha_0 = 7.93 \times 10^{-20}\ m^{-1}$ and $v_\infty = 150 \ km\ s ^{-1}$. The theoretical density profile is plotted on Figure $\ref{Densite}$, whereas the theoretical velocity profile is plotted on Figure $\ref{Vitesse}$ and can be compared to the observed one. Despite some small differences, the theoretical curve can be considered as a good approximation of the real one. It is interesting to note that to derive the theoretical rotation curve, no assumption had to be made on the global spatial distribution of matter, the knowledge of its distribution along the galactic plane was sufficient. On the contrary, the global spatial distribution of matter can be determined from the integration of Eq.\ $(\ref{AA22})$ and $(\ref{eqeq})$, starting from the knowledge of the fields along the galactic plane. Unfortunately, performing these integrations requires to define appropriate boundary conditions, and this is not an easy task.

It is important to highlight that according to Eq.\ $(\ref{yue})$, obtained by neglecting $k_z$, $\rho$ will be positive only if $\partial v/\partial r$ is also positive. This means that the assumption we made by neglecting $k_z$ is not valid for spiral galaxies presenting regions where $\partial v/\partial r < 0$. It is the action of $k_z$ that allows the slope of the rotation curve to present a negative value in some regions. This can be seen from Eq.\ $(\ref{rhonouveau})$ written as such:
\begin{equation}
	\frac{\partial v}{\partial r}\left(\frac{v}{r} + 2k_z\right) = 2\pi G \rho - 2k_z\frac{v}{r}\,.
\end{equation}
It has been justified in section $\ref{S4}$ that $k_z>0$ for a large part of the galaxy. Therefore, in the left hand side of this last relation, the term in parentheses is always positive. This means that the sign of $\partial v/\partial r$ is defined by the sign of the right hand side, and hence depends on how $2\pi G \rho$ and $2k_zv/r$ are related to each other. In regions where $2\pi G \rho > 2k_zv/r$, the derivative of the velocity is positive. On the other hand, in the regions where $2\pi G \rho < 2k_zv/r$, the derivative of the velocity is negative.

This hence justifies to improve the model we established above in order to describe the structure of more general spiral galaxies by taking into account a non-negligible $k_z$. This will be done as follows:
\begin{enumerate}
	\item To take into account the existence of the bulge, the density profile will be assumed to be described as the combination of two decreasing exponentials:
	\begin{equation}\label{rho2}
		\rho = \rho_0 e^{-\alpha_0 r} + \rho_1 e^{-\alpha_1 r}\,,
	\end{equation}
	where $\rho_1$ and $\alpha_1$ are positive constants, and are related to $\rho_0$ and $\alpha_0$ such that the first term of Eq.\ $(\ref{rho2})$ becomes dominant for larges radii, while the second term becomes dominant for smaller radii.
	
	\item Even small, $k_z$ is not exactly zero. As explained above, by analogy with electromagnetism, a spiral galaxy can be considered as an assembly of successive concentric current loops having different radii, in which the masses are the equivalent of the electric charges. In some places, these current loops develop gravitomagnetic fields which for some of them combine in a destructive manner, and for the others combine in a constructive manner. But the more we approach the center of the galaxy, the more these current loops combine in an overall constructive manner. So, if far from the center of the galaxy $k_z$ will indeed tend to zero, the more we approach its center, the more $k_z$ is expected to increase. We will hence assume that $k_z$ can be expressed as a decreasing exponential also:
	\begin{equation}\label{k1}
		k_z = k_0 e^{-\beta_0 r}\,,
	\end{equation}
	where $k_0$ and $\beta_0$ are positive constants.
\end{enumerate}

Inserting now the Eq.\ $(\ref{rho2})$ and $(\ref{k1})$ into Eq.\ $(\ref{rhonouveau})$, we find
\begin{eqnarray}\label{EEE}
	&&\frac{\partial v}{\partial r}\left(\frac{v}{r}+2k_0 e^{-\beta_0 r}\right) = 2\pi G \rho_1 e^{-\alpha_1 r} - 2k_0 \frac{v}{r}e^{-\beta_0 r}\nonumber
	\\
	&& + 2\pi G \rho_0 e^{-\alpha_0 r}\,.
\end{eqnarray}
We will show that Eq.\ $(\ref{EEE})$ is adequate to describe more complex rotation curves. In fact, depending on how the three terms of the right hand side are related to each other, several different rotation curves can be constructed. We will consider only one of them. We will fix $\rho_0$, $\alpha_0$, $\rho_1$, $\alpha_1$, $k_0$ and $\beta_0$ such that we may distinguish 3 regions, in which one of the three terms of the right hand side of Eq.\ $(\ref{EEE})$ is dominating the two others. For very small radii, the dominating term would be the one related to the bulge (i.e., the first term of the right hand side). According to Eq.\ $(\ref{EEE})$, the derivative of the velocity is hence positive, and the velocity increases for increasing radii. For intermediate radii, the term related to the general relativistic effects (i.e., the second term of the right hand side) becomes dominant. The derivative of the velocity becomes negative, and the velocity decreases. Finally, for larger radii, the dominating term is the one related to the disk region (i.e., the last term of the right hand side), and the velocity increases again to tend to its plateau.

\begin{figure}
	\centering
	\includegraphics[width=\linewidth]{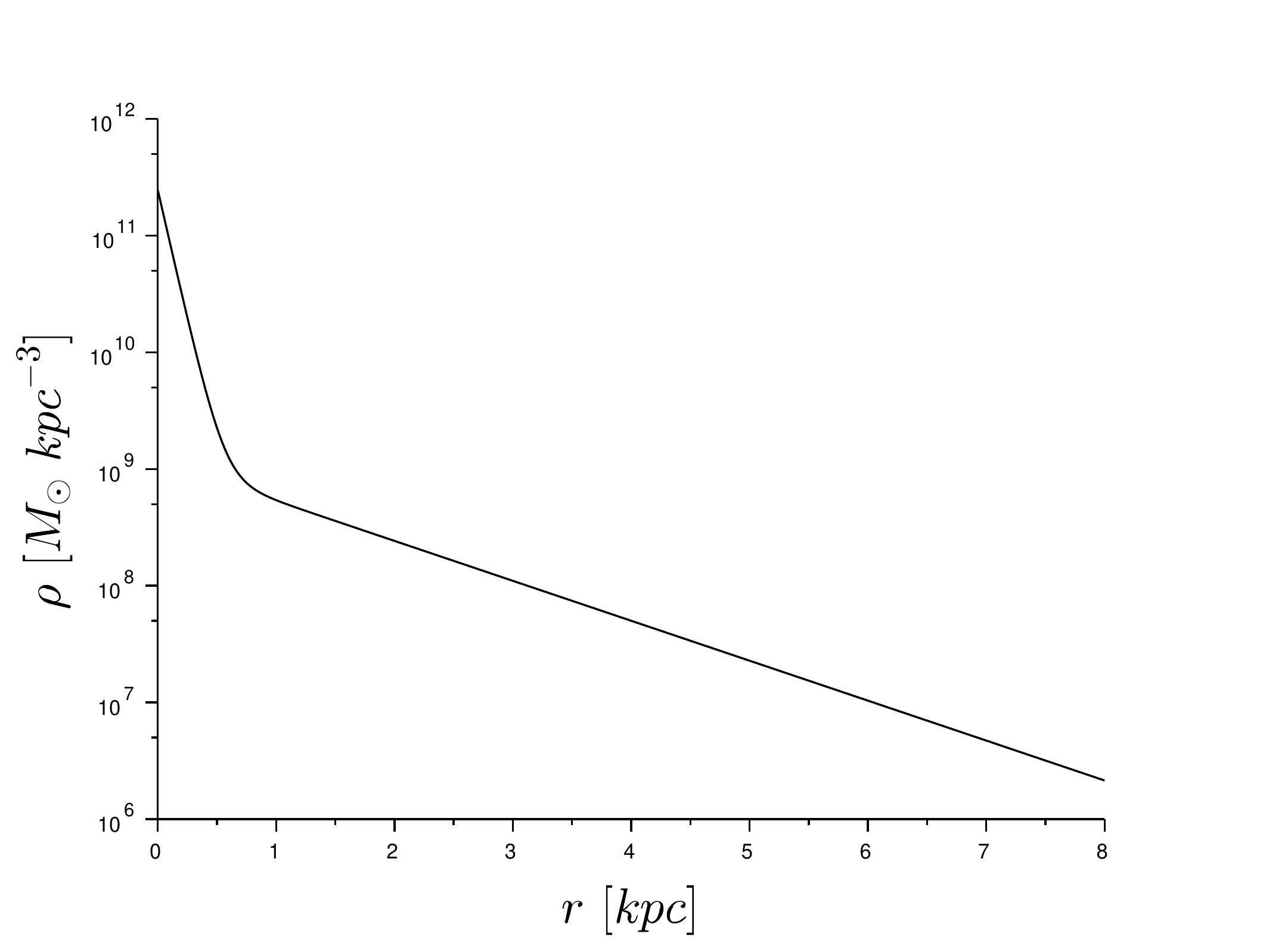}
	\captionsetup{width=0.5\textwidth}\caption{Theoretical density profile along the galactic plane of the Milky Way.}\label{Densite2}
\end{figure}

\begin{figure}
	\centering
	\includegraphics[width=\linewidth]{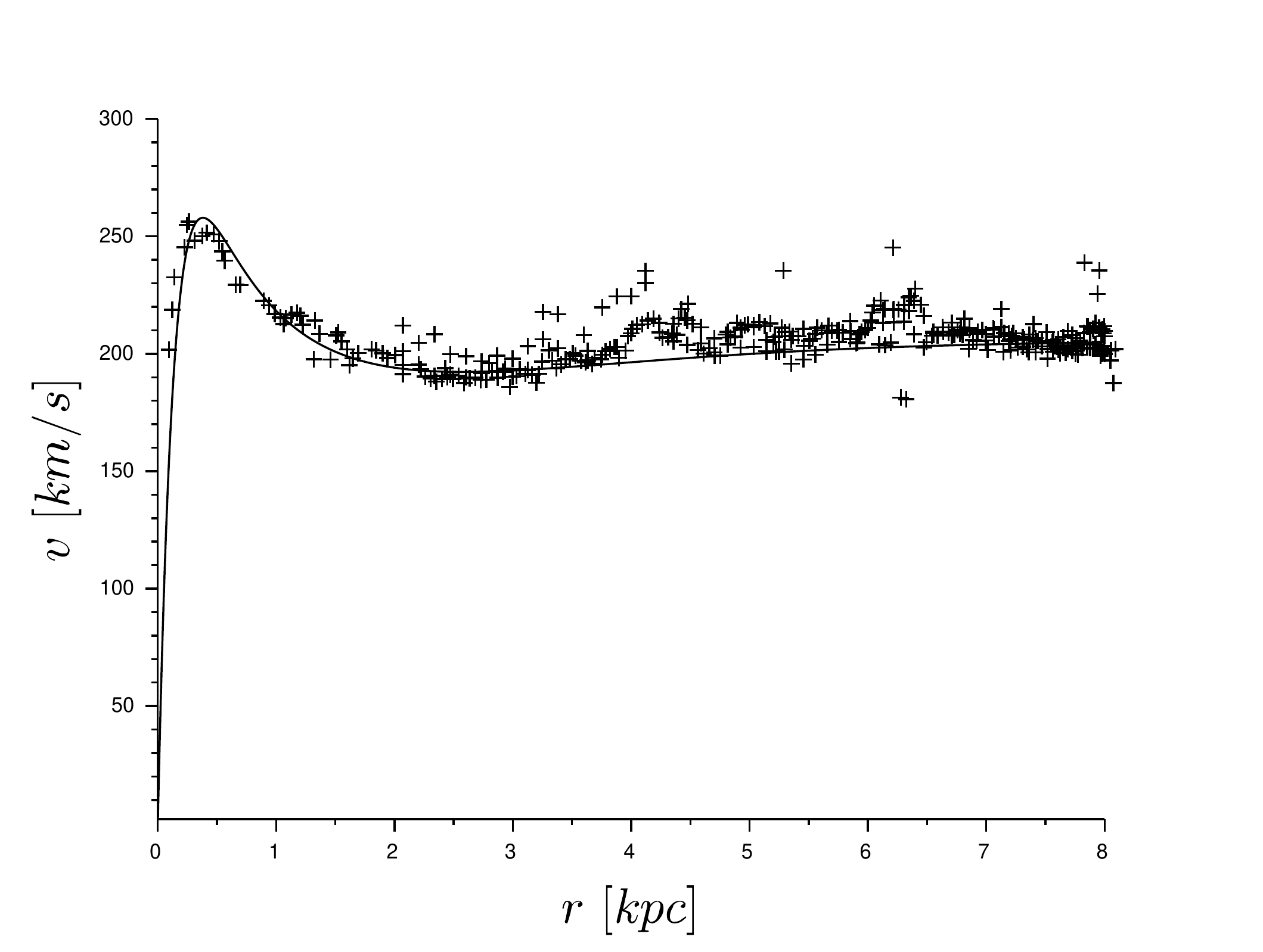}
	\captionsetup{width=0.5\textwidth}\caption{Rotation curve of the Milky Way. The solid line represents the prediction of the model, while de $+$ marks are observational data from \cite{Sofue}.}\label{Vitesse2}
\end{figure}

We consider the case of the Milky Way, and use the data from \cite{Sofue} for comparison. We then fix $\rho_0 = 7.91\times 10^{-20} \ kg\ m^{-3}$, $\alpha_0 = 2.55\times 10^{-20} \ m^{-1}$, $\rho_1 = 1.65\times 10^{-17} \ kg\ m^{-3}$, $\alpha_1 = 3.3 \times 10^{-19} \ m^{-1}$, $k_0 = 4.82\times 10^{-15} \ s^{-1}$ and $\beta_0 = 2.12\times 10^{-20} \ m^{-1}$. These values have been obtained by imposing the theoretical rotation curve to present some characteristics (typically a value or a derivative at some specific points, deduced from the observational data), and using an iterative process. The theoretical density profile is plotted on Figure $\ref{Densite2}$, whereas the theoretical velocity profile is plotted on Figure $\ref{Vitesse2}$ and can be compared to the observed one. Here also, the model is able to predict correctly the rotation curve, together with an adequate density profile along the galactic plane. As for the previous case, the knowledge of the density profile and of $k_z$ along the galactic plane was sufficient to deduce the rotation curve, in other words, the knowledge of the global spatial distribution of matter was not required. Also, it is interesting to note that, for this case, i.e. a spiral galaxy presenting a region with a decreasing velocity profile, the model suggests that general relativistic effects cannot be neglected, because they become dominant with respect to gravitational forces in this specific region.


\section{Conclusion}

We have established the stationary solution of a rotating pressureless fluid for the linearized approximation of the theory of general relativity to have a better understanding of the structure of spiral galaxies. On the basis of the positiveness of the density, we have demonstrated that this solution necessarily implies a constant velocity profile in the regions where the density and the gravitomagnetic fields are negligible, such as in the disk region of a spiral galaxy. This result is in contradiction with the Keplerian behavior which is commonly presented as the expected one in the disk region of spiral galaxies, at least if we suppose that the contribution of the masses outside of the bulge to the gravitational field are insignificant. General relativistic effects remain small in the disk region, but their consideration proved to be necessary to establish the constraint on the density, explaining the flatness of the rotation curve. Such constraint cannot be derived from a Newtonian approach. Dark matter is still required to explain how this flat rotation curve can be reached in the disk region. 

The model has also been applied on two specific spiral galaxies, to illustrate its ability to correctly predict their rotation curves. The results suggest that for some spiral galaxies, general relativistic effects cannot be neglected in regions close to the bulge, and should be taken into account to have a proper understanding of the rotation curve. This highlight the relevance of considering a general relativistic model to analyze the structure of spiral galaxies.


\bibliographystyle{unsrt}


\end{document}